\documentclass[a4paper, conference, 10pt]{IEEEtran}
\usepackage{graphicx}
\usepackage{times}
\usepackage{epsfig}
\usepackage{epstopdf}
\usepackage{latexsym}
\usepackage{amsmath}
\usepackage{textcomp}
\usepackage{enumerate}
\usepackage[noadjust]{cite}   
\usepackage{relsize}          
\usepackage{mathtools}
\usepackage{bbm}              
\usepackage{dsfont}           
\usepackage{mathbbol}
\usepackage[keeplastbox]{flushend}
\usepackage{subcaption}

\usepackage{fancyhdr}

\begin{document}

\title{Analysis of Analog Network Coding noise in Multiuser Cooperative Relaying for Spatially Correlated Environment}

\author{\IEEEauthorblockN{Sam~Darshi}\IEEEauthorblockA{Department of Electrical Engineering \\
IIT Ropar, Punjab\\
Email: sam@iitrpr.ac.in}\and \IEEEauthorblockN{Samar~Shailendra}\IEEEauthorblockA{TCS Research \& Innovation\\
Bangalore\\
Email: s.samar@tcs.com}}

\maketitle

\thispagestyle{fancy}

\begin{abstract}

Analog Network Coding (ANC) is proposed in literature to improve the network throughput by exploiting channel diversity.
In practical scenarios, due to the difference in channel characteristics an extra residual component, termed as ANC Noise, appears during the processing of the received signal. This ANC Noise component may suppress the ANC gain. None of the existing literature to our knowledge considers the effect of spatial correlation among channels on ANC noise. This paper develops a generic framework to investigate the effect of channel characteristics on ANC noise. We have modelled the channels as spatially correlated to take the (dis)similarity among them into account. Per node power constraint is also taken into consideration. In this work, we have characterized the behaviour of ANC noise and presented the results to analyze the network performance in terms of outage probability. Outcomes of our investigation show that spatial correlation among channels significantly affects the variance of ANC noise as well as the outage performance of the system. The proposed framework can provide better insights while selecting the system parameters in a correlated environment.

\end{abstract} 
\label{sec:abs}
\section{Introduction}


The use of Analog Network Coding (ANC) in combination with multiuser cooperative diversity is being proposed in the literature to achieve better throughput performance \cite{liu16}. In multiuser scenarios, a relay node may combine signals from more than one sources using ANC and help corresponding receivers to achieve better quality of service (QoS). Depending upon channel conditions, relay may employ either amplify and forward (AF) or decode and forward (DF) method for relaying purpose. However, it is shown in \cite{sharma2012} that the use of ANC in cooperative scenario may not always be prolific. During signal extraction process at the receiver, in addition to system noise, another undesired signal term known as ANC noise appears due to dissimilar channel characteristics which results in performance degradation \cite{mobi12}.


The amount of ANC noise greatly depends upon the relative characteristics of the channel between two nodes. Unlike the system noise (Additive White Gaussian Noise), which appears primarily due to receiver circuitry/processing, ANC noise depends upon the relative channel conditions. The deleterious effect of ANC noise increases with number of sources/sessions also \cite{sharma2012}.

To the best of our knowledge, the effects of channel correlation on ANC noise are not being addressed in the existing literature. In this paper, we have provided a generic framework to study the effect of difference in channel characteristics on ANC noise by taking  spatial correlation into consideration. The proposed framework also considers the per node power constraints. Effect of various system parameters is analyzed on ANC noise. Outage probability is calculated to observe the effect of ANC noise in the presence of channel correlation.

Rest of the paper is organized as follows : Section \ref{sec:sysdes} briefly describes the system model. In Section \ref{sec:third}, we presents the ANC noise analysis for uncorrelated (Section \ref{sec:third_uncor}) as well as for the correlated (Section \ref{sec:NC}) channels environment. The numerical results are discussed in Section \ref{sec:perf}. Finally, Section \ref{sec:conc} concludes the paper.

\label{sec:int}
\section{System Description}
\begin{figure} [ht!]
  \centering
  \includegraphics[height= 5.3 cm ,width=8.0cm]{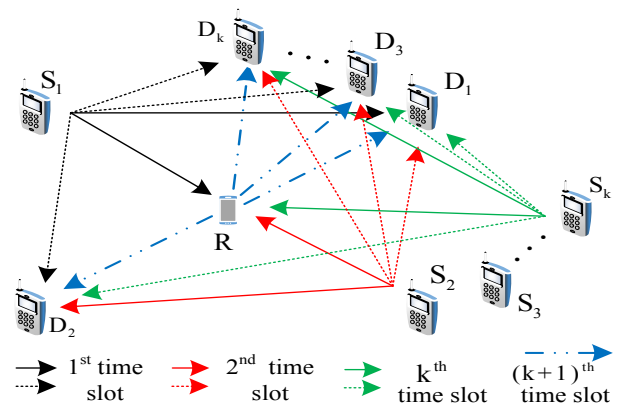}\\
  \caption{General network coded cooperative communication scenario.}\label{fig:nw1}
  \label{Fig:nw1}
\end{figure}

The network is assumed to be comprising of $k$ source - destination pairs. The communication among node pairs is assisted by a relay node (R) as shown in Fig \ref{Fig:nw1}.
\vspace{-4mm}
\vspace{-1.5mm}

Power distribution among the network nodes is based on the total network power constraint \cite{sad08}. If $P_\text{Tot}$ is the total power constraint of the network (with $k$ sources) under consideration as in Fig. \ref{fig:nw1}, the power distribution among nodes can be written as \( \sum_{{S_i}} {{P_{{S_i}}}\, + \,} \,{P_R}\, = \,{P_\text{Tot}},\) where $S=\{S_i\mid 1\leq i\leq k\}$ is the set of all sources. $P_{S_i}$, $P_R$ are the transmission powers of $i^\text{th}$ source node, and the relay node respectively.
Following power distribution model is assumed for the scenario in hand.
\begin{equation}\label{dis}
\begin{gathered}
  {P_{{S_i}}}\, = \,{\psi _i}\,{P_\text{Tot}}\,\,, \hfill \\
  {P_R}\,\,\, = \, {\left(1- {\sum\limits_{{i}} {\psi _i}} \right){P_\text{Tot}} } \,,\,\,\,\,\,\,\, i = 1,2,..k \hfill \\
\end{gathered}
\end{equation}
\vspace{-1mm}

\noindent where $\psi_i\in \{(0,1] \, \vert \,  \sum_{i} {\psi_i} \leq 1 , \forall i\}$; here equality condition leads to non-cooperative communication as $P_R=0$.  Note that introduction of the above power constraint provides us a unique opportunity to use the proposed framework to analyze the networks containing power constraint devices. However, this work does not deal with the optimal power distribution among the nodes.

In the network under consideration (Fig. \ref{fig:nw1}), the overall communication process is divided into $(k+1)$ time slots. The $k$ sources transmit in $k$ time slots sequentially where every transmission is received by all the destination nodes (along with the desired one) as well as the relay node i.e. In first time slot, source 1 ($S_1$) transmits, which is received by all destination nodes ($D_1, D_2,...,D_k$) including the relay node ($R$). In second time slot, source 2 ($S_2$) transmits, which is again received by all the destination nodes ($D_1, D_2,...,D_k$) including $R$ and so on for subsequent time slots i.e. till $k^\text{th}$ time slot. Once, all the sources finish their transmission, $R$ combines all the signals received during $k$ transmission slots using ANC and transmits this combined signal using AF method \cite{patel06} to all destination nodes in $(k+1)^\text{th}$ slot. It can be noted that while source $S_j$ transmits its signal, $D_j$ receives this as the desired signal whereas, other receivers overhear this as an undesired signal. However, this undesired signal is utilized in decoding the network coded signal received in  $(k+1)^\text{th}$ slot, as explained in Section \ref{sec:third}.

%

In order to achieve diversity, all receiver nodes try to extract another copy of the respective desired signal from the combined network coded signal, received in  $(k+1)^\text{th}$ slot.
In process of extracting the desired signal from the combined  network coded signal, received from node $R$, each destination node subtracts the overheard signals, received in  previous time slots. After successful extraction of second copy of the desired signal, any of the combining techniques \cite{gold05} can be applied to get better performance.

The extraction process at each destination node, involves the subtraction of signals, received at the destination node over different paths. However, due to the difference in the channel conditions it results in a residual undesired term which contributes to the noise at the receiver and in turn degrades the performance. This residual term is known as the analog network coding (ANC) noise \cite{sharma2012} or NC noise \cite{mobi12}. This term also depends upon the spatial correlation between the channels. To the best of our knowledge such a spatial correlation has not been considered in any of the earlier work. We have studied the effect of such correlation in this paper and results are presented in the subsequent sections. For our studies, we have considered that all destination nodes are in transmission range of all source nodes. Each node consists of single antenna, operates in half duplex mode and all channels are assumed to be frequency flat.

\label{sec:sysdes}
\section{ANC Noise Analysis} \label{sec:third}

In this section, we first discuss the modelling of ANC noise for uncorrelated channel scenario (Section \ref{sec:third_uncor}) followed by its extension for the correlated environment in the subsequent subsection (Section \ref{sec:NC}).

\subsection{\textbf{ANC Noise: Uncorrelated Environment}}
\label{sec:third_uncor}
\subsubsection{\textbf{Modelling for two sources and a relay node scenario}}
\label{sec:2SR}
\begin{figure} [ht!]
  \centering
  \includegraphics[height= 4.0 cm ,width=8.3cm]{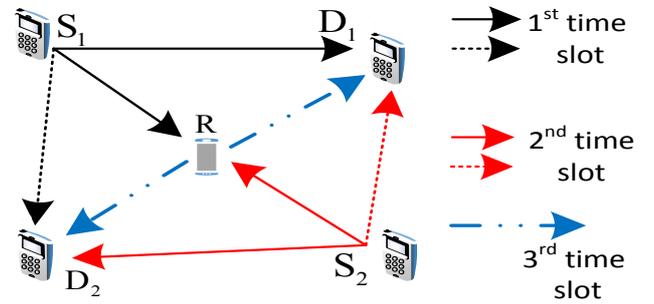}\\
  \caption{Network coded cooperative scenario with two source nodes and a relay node.}\label{fig:nw2}
  \label{Fig:nw2}
\end{figure}

To explain the ANC noise in uncorrelated environment, for simplicity, we have considered a network coded cooperative scenario with $k = 2$ and one relay node in the network (Fig. \ref{fig:nw2}). However, the same analysis can be easily extended for any number of source nodes and the results are produced later in this subsection.

\vspace{-3mm}

\vspace{-1mm}

For ease of explanation, all the transmissions and receptions are listed below assuming that receiver node has complete CSI. \\
In time slot $1$, signals received  at $D_1$, $D_2$ and $R$ can be represented as
\vspace{-3.0mm}
\begin{equation}\label{eq1}
 Y_{{S_1}{D_j}} =\,\, d^{-\alpha/2}_{{S_1}{D_j}}\,\,\, h_{{S_1}{D_j}}\, x_1 \,\, + \,\, n_{D_j}, \quad j=1,2
\end{equation}

\vspace{-2.3mm}

\begin{equation}\label{eq2}
 Y_{{S_1}R} =\,\, d^{-\alpha/2}_{{S_1}{R}}\,\,\, h_{{S_1}R}\, x_1 \,\,\,\, + \,\,\, n_R
\end{equation}

%

\noindent where $Y_{ab}$, $d_{ab}$, $h_{ab}$ and $x_1$ are the signal received at node $b$ from node $a$, distance between  node $a$ and $b$, channel coefficient$^{1}$  between node $a$ and $b$ and signal transmitted in slot $1$, respectively. Symbol $\alpha$ denotes the pathloss exponent and $n_a$ is AWGN at node $a$ with variance $\sigma^2_a$.

\footnotetext[1] {The effect of distance can also be accommodated in $h$ as shown in \cite{sharma2012}.}

Similarly, in time slot $2$, signals received  at $D_1$, $D_2$ and $R$ can be represented as
\vspace{-1mm}
\begin{equation}\label{eq4}
 Y_{{S_2}{D_j}} = \,\, d^{-\alpha/2}_{{S_2}{D_j}}\,\,\, h_{{S_2}{D_j}}\, x_2 \,\, + \,\, n_{D_j}, \quad j=1,2
\end{equation}

\vspace{-2.2mm}
\begin{equation}\label{eq5}
 Y_{{S_2}R} = \,\, d^{-\alpha/2}_{{S_2}{R}}\,\,\, h_{{S_2}R}\, x_2 \,\,\,\,\, + \,\,\, n_R
\end{equation}


\noindent Signal received by destinations in time slot $3$ (transmitted by $R$),
\vspace{-1mm}
\begin{equation}\label{eq7}
 Y_{{R}{D_j}} = \,\, d^{-\alpha/2}_{{R}{D_j}}\,\,\, h_{{R}{D_j}}\, X_R \,\, + \,\, n_{D_j}, \quad j=1,2
\end{equation}

\noindent where $X_R$ denotes the analog network coded signal transmitted by $R$ using AF method.
Further, $X_R$ can be written as 
$ X_{R} =  A_f\, [Y_{{S_1}{R}}   +  Y_{{S_2}{R}}]$,
where $A_f$ is the amplification factor used by relay for AF purpose \cite{sharma2012,lane04}. The value of $A_f$
can be written as 
\begin{equation}\label{eq10}
{A_f}\, = \,\sqrt {\frac{{{P_R}}}{\displaystyle {\sum\nolimits_{m = 1}^2 {\left( {\frac{\displaystyle {{P_{{S_m}}}\,{{\left| {{h_{{S_m}R}}} \right|}^2}}}{\displaystyle {d_{{S_m}R}^\alpha }}} \right)\, + \,\,2\sigma _R^2} }}}.
\end{equation}

Once, the transmission from $R$ is completed, all destination nodes initiate their extraction process for second copy of the respective desired signal. Without loss of generality,
let's choose $D_1$ as the node of interest. The signals received by $D_1$ in time slot $1$ and $2$ are given by (\ref{eq1}) and (\ref{eq4}), respectively. At node $D_1$, the combined signal in slot $3$  can also be written as,
\begin{equation}\label{eq12}
Y_{{R}{D_1}} = \,\, d^{-\alpha/2}_{{R}{D_1}}\,\,\, h_{{R}{D_1}}\, A_f\, [Y_{{S_1}{R}}   +  Y_{{S_2}{R}}] \,\, + \,\, n_{D_1}
\end{equation}

\noindent Substituting $Y_{{S_2}{R}}$ from (\ref{eq5}) yields
\begin{equation}\label{eq13}
 Y_{{R}{D_1}}= d^{-\alpha/2}_{{R}{D_1}}\, h_{{R}{D_1}}\, A_f\, [Y_{{S_1}{R}}+\, d^{-\alpha/2}_{{S_2}{R}}\, h_{{S_2}R}\, x_2 \, + \, n_R] \,+ \, n_{D_1}.
\end{equation}

Assuming that channel gains and amplification factors are known apriori at $D_1$, reconstructed signal can be obtained by performing following operation at $D_1$ as

\vspace{-3mm}

\begin{equation}\label{eq14}
{{\mathop Y\limits^ \sim}_{R{D_1}}}\, = \,\,{Y_{R{D_1}}}\, - \,\,\frac{{{A_f}{\, d^{-\alpha/2}_{{R}{D_1}}\,}{h_{R{D_1}}}{\, d^{-\alpha/2}_{{S_2}{R}}\,} {h_{{S_2}R}}\,}}{{\, d^{-\alpha/2}_{{S_2}{D_1}}\,}{{h_{{S_2}{D_1}}}}}\,{Y_{{S_2}{D_1}}}
\end{equation}

\noindent Using (\ref{eq4}) and (\ref{eq12}) in (\ref{eq14}) yields,



\vspace{-3mm}

\begin{equation} \label{eq15}
\begin{multlined}
\,{{\mathop Y\limits^ \sim}_{R{D_1}}}\, = \,\underbrace {{A_f}{\, d^{-\alpha/2}_{{R}{D_1}}\,} {h_{R{D_1}}}{Y_{{S_1}R}}}_{{1^{st}}\,\text{component}}\,\,\,\, + \underbrace {{n_{{D_1}}}}_{{2^{nd}}\,\text{component}}\, \vspace{3mm} \, \\
\,\,\,\,\,\,\,\,\,\,\,\,\,\,\,\,\,\,\,\,\,\,\,\,\,\,\,\,\,\,\,\,\,\,\,\,\,\,\,\,\,\,\,\,\,\,\,\,\,\,\,\, \hspace{-33mm} + \,\underbrace {{A_f}{\, d^{-\alpha/2}_{{R}{D_1}}\,} {h_{R{D_1}}}{n_R}\, - \,\frac{{{A_f}{{\, d^{\alpha/2}_{{S_2}{D_1}}\,}} {h_{R{D_1}}}{h_{{S_2}R}}\,}}{{\, d^{\alpha/2}_{{R}{D_1}}\,}{\, d^{\alpha/2}_{{S_2}{R}}\,}{{h_{{S_2}{D_1}}}}}\,{n_{{D_1}}}}_{{3^{rd}}\,\text{component}}
\end{multlined}
\end{equation}


From (\ref{eq15}), it can be observed that the second copy of desired signal at $D_1$, obtained by performing reconstruction using signals from slot $2$ and $3$, consists of 3 components. These components correspond to the desired signal, background noise and additional undesired signal term, respectively. The third undesired term is due to ANC noise.

This undesired signal term results due to the involvement of the channel parameters in reconstruction process. In order to extract the desired signal from the network coded signal received from $R$, node $D_1$ has to subtract the overheard signal in slot 2. Since, the direct signal from $S_2$ involves channel coefficient $h_{{S_2}D_1}$ and the signal via $R$ involves channel coefficients  $h_{{S_2}R}$ and $h_{R{D_1}}$, the combined effect of $h_{{S_2}R}$ and $h_{R{D_1}}$ may not be same as $h_{{S_2}D_1}$. Therefore, subtraction process results in a residual term i.e. the ANC noise.

The overall noise \cite{sharma2012} in the system can be written as
\begin{equation}\label{NC}
\begin{multlined}
\mathbb{N}_{D_1}^{NC}\, = \,\,{n_{{D_1}}}\, + \,\,{A_f}{{\, d^{-\alpha/2}_{{R}{D_1}}\,}} {h_{R{D_1}}}{n_R}\, -
\vspace{3mm} \, \\
\,\,\,\,\,\,\,\,\,\,\,\,\,\,\,\,\,\,\,\,\,\,\,\,\,\,\,\,\,\,\,\,\, \hspace{16mm}\, \frac{{{A_f}{{\, d^{\alpha/2}_{{S_2}{D_1}}\,}} {h_{R{D_1}}}{h_{{S_2}R}}\,}}{{\, d^{\alpha/2}_{{R}{D_1}}\,}{\, d^{\alpha/2}_{{S_2}{R}}\,}{{h_{{S_2}{D_1}}}}}     \,{n_{{D_1}}}\
\end{multlined}
\end{equation}

\noindent while the variance of total noise term can be found as
\begin{equation}\label{var}
\begin{multlined}
\sigma _{\mathbb{N}_{{D_1}}^{NC}}^2\, = \,\,\sigma _{{D_1}}^2 + \,\sigma _R^2\,{\left( {{A_f}{{\, d^{-\alpha/2}_{{R}{D_1}}\,}} {h_{R{D_1}}}} \right)^2}\, + \, \vspace{1.5mm} \, \\  \,\,\,
\hspace{28mm} \,\sigma _{{D_1}}^2  {\left( \frac{{{A_f}{{\, d^{\alpha/2}_{{S_2}{D_1}}\,}} {h_{R{D_1}}}{h_{{S_2}R}}\,}}{{\, d^{\alpha/2}_{{R}{D_1}}\,}{\, d^{\alpha/2}_{{S_2}{R}}\,}{{h_{{S_2}{D_1}}}}} \right)^2}
\end{multlined}
\end{equation}


\subsubsection{\textbf{Modelling for general case of K sources and a relay node scenario}}
\label{sec:NSR}
After developing the  statistics for a specific value of $k$, here we discuss the case of arbitrary number of source-destination pairs $k=K$ assisted by a relay node. Network operates in the similar fashion as described earlier. For $K$ source-destination pairs, total $(K+1)$  time slots are required. Based on the expression obtained in (\ref{var}), it is easy to extend it for a the general case of $K$ source-destination pairs. Therefore variance of overall noise for this general case may be written as follows

\vspace{-5mm}

\begin{equation}\label{general}
\begin{multlined}
\sigma _{\mathbb{N}_{{D_i}}^{NC}}^2\, = \,\,\sigma _{{D_i}}^2 + (K-1)\,\sigma _R^2\,{\left( {{A_f}{{\, d^{-\alpha/2}_{{R}{D_i}}\,}} {h_{R{D_i}}}} \right)^2}\, + \, \vspace{1.5mm} \, \\  \,\,\,
\hspace{12mm} \,\sigma _{{D_i}}^2  \hspace{-3mm} \sum\limits_{\begin{array}{*{20}{c}}
{ \forall \,\, {S_j} }\\
{{S_j} \ne {S_i}}
\end{array}} {\left( \frac{{{A_f}{{\, d^{\alpha/2}_{{S_j}{D_i}}\,}} {h_{R{D_i}}}{h_{{S_j}R}}\,}}{{\, d^{\alpha/2}_{{R}{D_i}}\,}{\, d^{\alpha/2}_{{S_j}{R}}\,}{{h_{{S_j}{D_i}}}}} \right)^2}
\end{multlined}
\end{equation}

\noindent where amplification factor for this general case can be obtained by generalizing (\ref{eq10}) as

\vspace{-4mm}

\begin{equation}\label{Agen}
{A_f}\, = \,\sqrt {\frac{{{P_R}}}{\displaystyle {\sum\nolimits_{m = 1}^K {\left( {\frac{\displaystyle {{P_{{S_m}}}\,{{\left| {{h_{{S_m}R}}} \right|}^2}}}{\displaystyle {d_{{S_m}R}^\alpha }}} \right)\, + \,K\,\sigma _R^2} }}}
\end{equation}

From (\ref{general}), it may be noted that the variance of ANC noise as well as the overall noise increases as the number of participating source-destination pair increases. This is intuitively supported by the observation that with more number of transmission sessions, there is an increase in overheard signals at each destination. The imperfect cancellation of these signals due to the difference in channel characteristics results in increase of the ANC noise.

%

\subsection{\textbf{ANC Noise : Correlated Environment}} \label{sec:NC}
As discussed in Section \ref{sec:third_uncor}, the differences in channel characteristics (dissimilarity) lead to residual additional term (ANC noise)  which degrades the system performance. This section investigates the generation of ANC noise term in more realistic environment by considering spatially correlated channel environment.

\begin{figure} [h]
  \centering
  \includegraphics[height= 2.8 cm ,width=7.8cm]{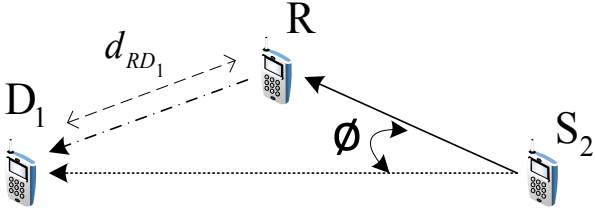}\\
  \caption{Spatially correlated scenario as a function of $\phi$. }\label{fig:corr}
  \label{Fig:pp}
\end{figure}

The extent of spatial correlation depends upon the relative position of nodes as well as the environmental factors. Since, the dissimilarity among channel characteristics is one of the main reasons for the occurrence of ANC noise term, the amount of ANC noise will be a function of channel dissimilarities. To take the measure of channel dissimilarity into account, channels between nodes are assumed to be spatially correlated. Two channels having high spatial correlation factor show very similar effect on signals, while channels with low correlation factor affect the signals quite differently.
Correlation between channels can also be expressed in terms of their overlap.
For example, in Fig. \ref{Fig:pp}, the overlap between direct path ($h_{{S_2}{D_1}}$) and relay path  ($h_{{S_2}R}$, $h_{R{D_1}}$)
may be expressed as a function of $\phi$ or $d_{R{D_1}}$.
\vspace{-3.5mm}
\vspace{-3.5mm}
For small values of $d_{R{D_1}}$ i.e. smaller $\phi$, the effect of direct path as well as the paths via $R$ on the signal is quite similar due to highly correlated channels (large overlap), the amount of residual ANC noise component ($Y_{R{D_1}}-Y_{{S_2}{D_1}}$) in signal remains very small.
On the contrary, for high values of $d_{R{D_1}}$ or $\phi$, the received signals at the receiver via the direct path and the relay path are significantly different (due to low overlap) which leads to much larger ANC noise at the receiver. This behaviour of ANC noise is further presented in Section \ref{sec:result}.


Various models are available in literature to calculate the coefficient of spatial correlation as a function of distance and angle between nodes in overlapping channel scenario \cite{wicz10, yam06}. In the literature, all channels are assumed to be spatially correlated among themselves defined by the specific correlation factors.

A careful observation reveals that while considering spatial correlation among channels, not all channels between the nodes affect the ANC noise. The channels that contribute to the ANC noise may be listed as follows; i) channels between source nodes to all undesired destination nodes ($h_{{S_i}D_j}, i\neq j$), ii) channels between source nodes to relay node ($h_{{S_i}R}$), iii) channels between relay node to destination nodes ($h_{{R}D_j}$). For a network consists of $K$ source-destination pairs, the number of channels ($N_{ch}$) affecting the ANC noise will be \( N_{ch}\, = \, K(K-1)\, +\, K \,+ \,K\, = \, K(K+1)\, \), where each term corresponds to three categories of channels as described above, namely $h_{{S_i}D_j}$, $h_{{S_i}R}$ and $h_{{R}D_j}$, $\forall i,j ; i\neq j$, respectively. Assuming that ANC noise due to each undesired source is independent for each destination node, these spatially correlated channels can be represented in the form of a correlation matrix ($\Gamma_{ij}$). For $i^{th}$ source and $j^{th}$ destination node ($i \neq j$), $\Gamma_{ij}$ can be given by,


\vspace{-4mm}

\begin{equation}\label{11}
\Gamma_{ij} \, = \,\left[ {\begin{array}{*{20}{c}}
 {\rho _{S_iD_j,S_iD_j}}&{{\rho _{S_iD_j,S_iR}}}& {{\rho _{S_iD_j,RD_j}}}\\
{{\rho _{S_iR,S_iD_j}}}& {\rho _{S_iR,S_iR}} & {{\rho _{S_iR,RD_j}}}\\
 {{\rho _{RD_j,S_iD_j}}}&{{\rho _{RD_j,S_iR}}}& {\rho _{RD_j,RD_j}}
\end{array}} \right]
\end{equation}

\noindent where  $1\leq i, j\leq K ; i\neq j$ and $\rho_{{ch_a},{ch_b}}$ is correlation  coefficient of the channel $ch_a$ and $ch_b$.
Network performance under the effect of correlation among other channels can be found in literature \cite{zha08}.

%
%

\label{sec:analy}
\section{RESULTS AND DISCUSSION } \label{sec:result}

In this section, we present the numerical study of the effect of spatial correlated channels over ANC noise discussed in Section \ref{sec:NC}. To verify the analysis, extensive simulations are being performed using MATLAB. We have simulated the two source-destination pair scenario $(k=2)$ as shown in Fig. \ref{fig:nw2}. The position of $R$ is kept symmetrical w.r.t. source nodes. All channels are assumed to be $\mathcal{CN}(0,1)$. For the simulation purposes, same noise variance is considered at each node, i.e. $\sigma^2_{D_1} = \sigma^2_R = \sigma^2_0$, total variance in (\ref{var}) is simplified as in (\ref{var2}).

\vspace{-4mm}

\begin{equation}\label{var2}
\begin{gathered}
\sigma _{\mathbb{N}_{{D_1}}^{NC}}^2\, = \,\sigma _{0}^2 \left( 1 + {\left( {{A_f}{{\, d^{-\alpha/2}_{{R}{D_1}}\,}}{h_{R{D_1}}}} \right)^2}\, + \right.\, \hfill \\
  \,\,\,\,\,\,\,\,\,\,\,\,\,\,\left.
\, \hspace{28mm} {\left( {\frac{{{A_f}{\, d^{\alpha/2}_{{S_2}{D_1}}\,} {h_{{S_2}R}}{h_{R{D_1}}}}}{{{{{\,d^{\alpha/2}_{{R}{D_1}}\,}{\, d^{\alpha/2}_{{S_2}{R}}\,}{{h_{{S_2}{D_1}}}}}}}}} \right)^2} \right) \hfill \\
\end{gathered}
\end{equation}

\noindent Other parameters used for the simulations are provided in Table \ref{table:simu}.


\begin{table} [ht!]
  \centering
  \caption{Simulation parameters}
  \begin{tabular}{|p{3.5cm}|p{2.7cm}|}
  \hline
  {\footnotesize {\fontsize{10}{10}\selectfont {\emph{\textbf{Parameter}}}}   } & {\footnotesize {\fontsize{10}{10}\selectfont {\emph{\textbf{Value}}}}   }  \\
  \hline
  \hline
  {\footnotesize Number of source node $(k)$ } &  {\footnotesize 2} \\
  \hline
  {\footnotesize Bandwidth} & {\footnotesize 22 MHz}  \\
  \hline
  {\footnotesize Path loss exponent ($\alpha$) } & {\footnotesize 4}  \\
  \hline
  {\footnotesize $d_{{S_1}{D_1}}$, $d_{{S_2}{D_2}}$ } & {\footnotesize 400 - 3700 m}  \\
  \hline
  {\footnotesize $d_{{S_2}{D_1}}$, $d_{{S_1}{D_2}}$ } & {\footnotesize 200 - 3500 m }  \\
  \hline
  {\footnotesize $d_{{S_1}{R}}$, $d_{{S_2}{R}}$, $d_{{R}{D_1}}$  } & {\footnotesize 223.60 - 2546.6 m }  \\
  \hline
  {\footnotesize Outage threshold ($\beta$) } & {\footnotesize $0 \, dB$  }  \\
  \hline
  {\footnotesize Available power ($P_\text{Tot}$)} & {\footnotesize 2 }  \\
  \hline
  {\footnotesize  $\psi_i$} & {\footnotesize  3/8 }  \\
  \hline
  \end{tabular}
  \label{table:simu}
\end{table}


For simulation purpose, without loss of generality, we take $\psi_i = (3/8)$ and $P_\text{Tot} = 2$. Each source node uses the same transmission power. However, framework is applicable to any value of $\psi_i$.

For the case under consideration in simulation $k = 2$, only three channels, namely $h_{{S_2}D_1}$, $h_{{S_2}R}$, $h_{RD_1}$, and other three channels, namely $h_{{S_1}D_2}$, $h_{{S_1}R}$ and $h_{RD_2}$ affect the ANC noise at destination nodes $D_1$ and $D_2$, respectively. Therefore, correlation matrix defined in (\ref{11}) may be written for $(S_2,D_1)$ pair as follows,

\vspace{-3mm}

\begin{equation}\label{simu_rho}
\Gamma_{21} \, = \,\left[ {\begin{array}{*{20}{c}}
 {\rho _{S_2D_1,S_2D_1}}&{{\rho _{S_2D_1,S_2R}}}& {{\rho _{S_2D_1,RD_1}}}\\
{{\rho _{S_2R,S_2D_1}}}& {\rho _{S_2R,S_2R}} & {{\rho _{S_2R,RD_1}}}\\
 {{\rho _{RD_1,S_2D_1}}}&{{\rho _{RD_1,S_2R}}}& {\rho _{RD_1,RD_1}}
\end{array}} \right]
\end{equation}

\noindent where $\rho_{{ch_a},{ch_b}}$ = 1 for $a=b$. Similarly, a correlation matrix $\Gamma_{12}$ for pair $(S_1,D_2)$ can also be defined.

For evaluation of system performance, Selection Combining \cite{gold05} is considered at the destination nodes. System behaviour is observed w.r.t. spatial correlation among channel coefficients and distance between various node pairs in terms of variance of ANC noise and outage probability. Results for ANC noise are presented first. For the ease of explanation, results are presented in two parts: for small distances and large distances, respectively. Large distances may not be practical for networks having battery operated terminals, but the study gives insights for other resource full network scenarios.
\vspace{-2mm}
\begin{figure} [ht!]
  \centering
  \includegraphics[height= 5.5cm ,width=8.0cm]{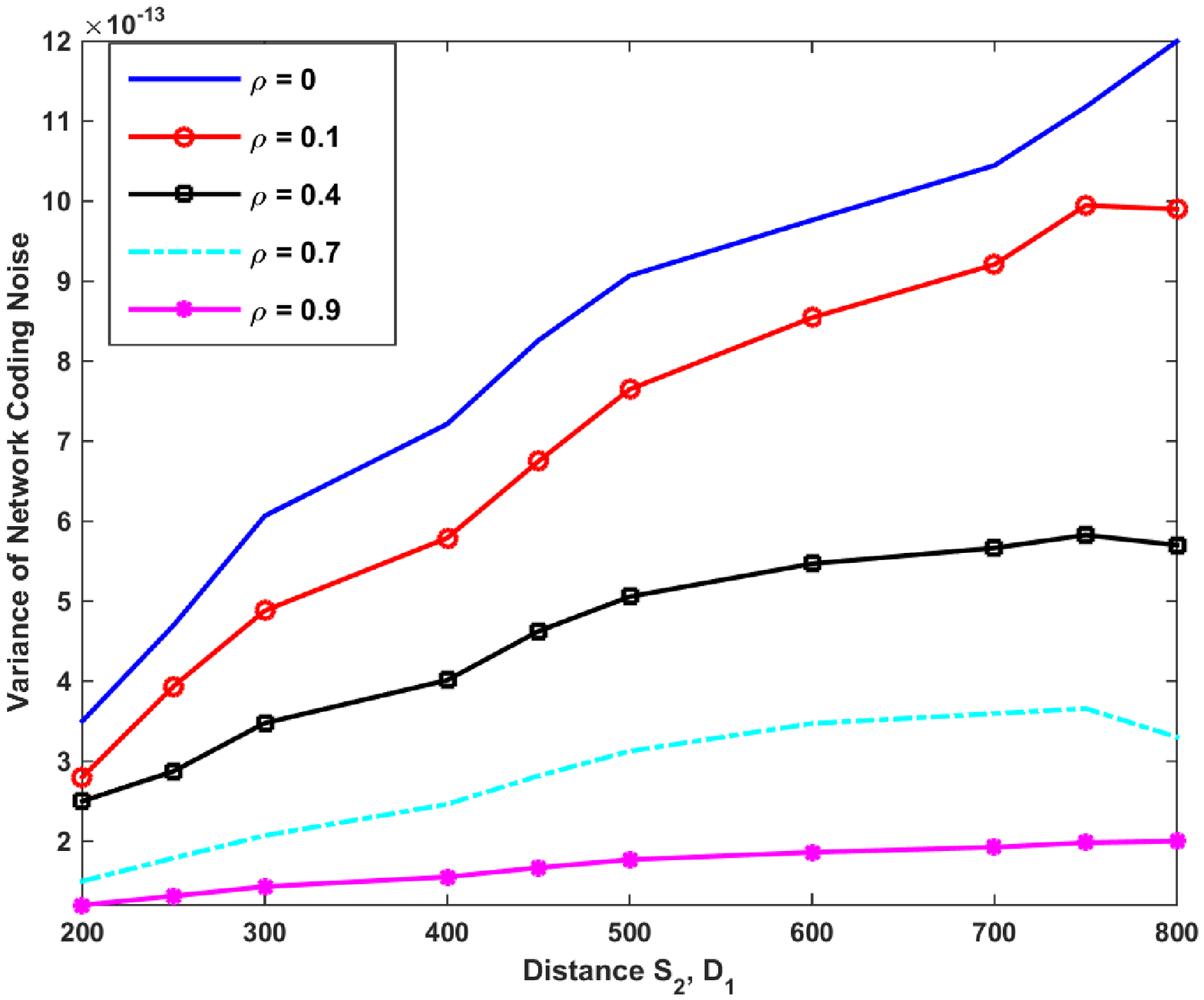}\\
  \caption{Variance of ANC noise with small distance for various correlation factors. }
  \label{Fig:res4}
\end{figure}
\vspace{-2mm}
Variation of ANC noise with distance among various nodes for different correlation factors is shown in Fig. \ref{Fig:res4}. Since, each element of various distance vectors shows one to one correspondence with each other, therefore for illustration purpose $d_{{S_2}{D_1}}$  is taken as a reference in Fig. \ref{Fig:res4}. For the given network configuration,
it can  be observed that the variance of ANC noise ($\sigma _{\mathbb{N}_{{D_1}}^{NC}}^2\,$) (given by (\ref{var2})) increases with an increase in distance between source ($S_2$) and other (than desired) receiver ($D_1$).
At the same time, $\sigma _{\mathbb{N}_{{D_1}}^{NC}}^2\,$ decreases with the increase in correlation ($\rho$) among channels  involved. One of the main reasons of this behavior is: as the correlation between the direct channel and the relay channel increases, the undesired components in the two received signals cancel out each other resulting in smaller ANC noise component at the receiver.
\vspace{-2mm}
\begin{figure} [ht!]
  \centering
  \includegraphics[height= 5.5cm ,width=8.0cm]{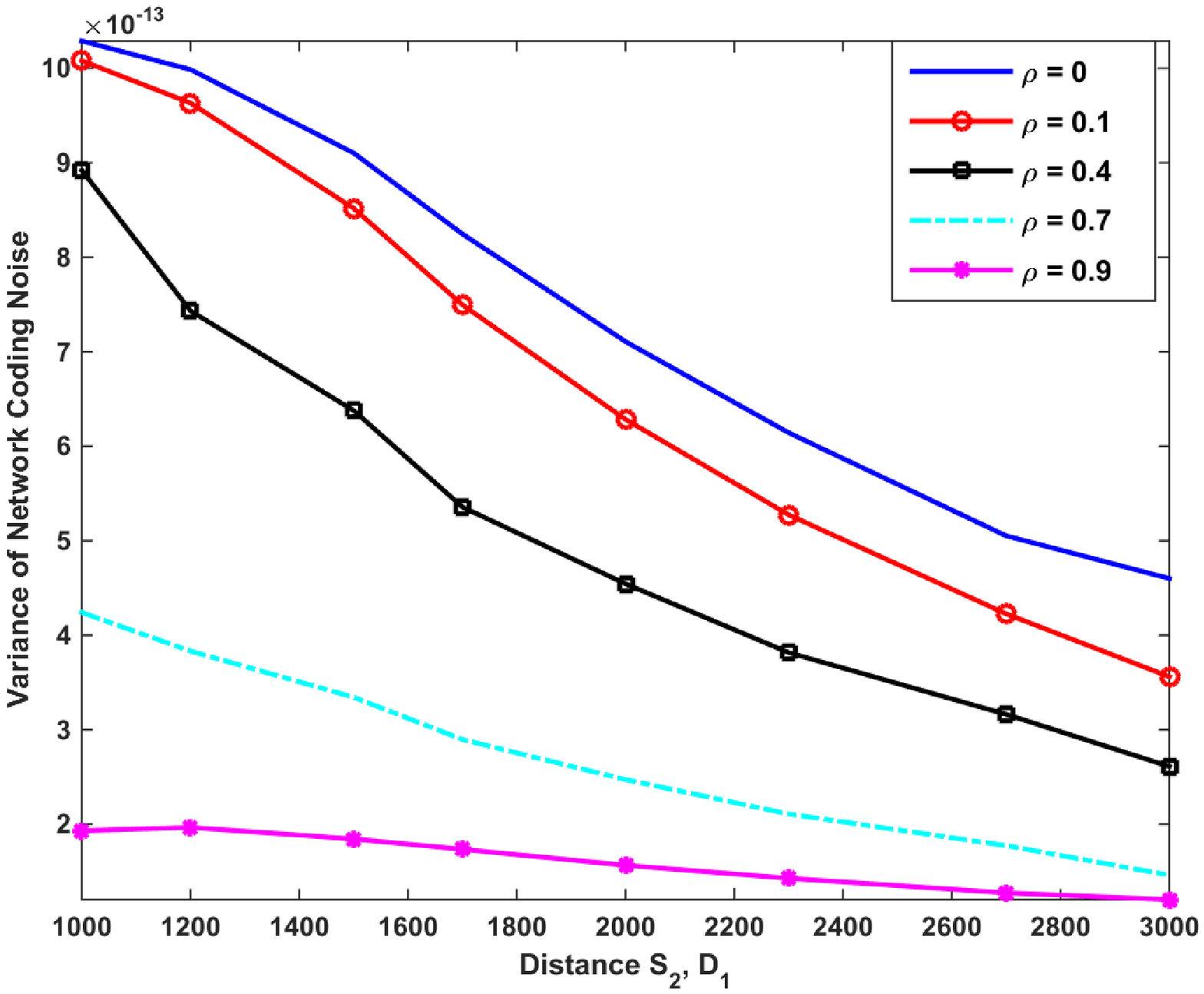}\\
  \caption{Variance of ANC noise with large distance for various correlation factors.}
  \label{Fig:res5}
\end{figure}
\vspace{-2mm}

However, a complete opposite demeanor of ANC noise variance can be observed in Fig. \ref{Fig:res5}. After a \emph{critical distance} ($\approx 900 - 1000 $ in this case), increase in distance among nodes, leads to reduction in ANC noise variance while the correlation shows the same characteristics as in Fig. \ref{Fig:res4}.
\vspace{-2mm}
\begin{figure} [ht!]
  \centering
  \includegraphics[height= 5.5cm ,width=8.0cm]{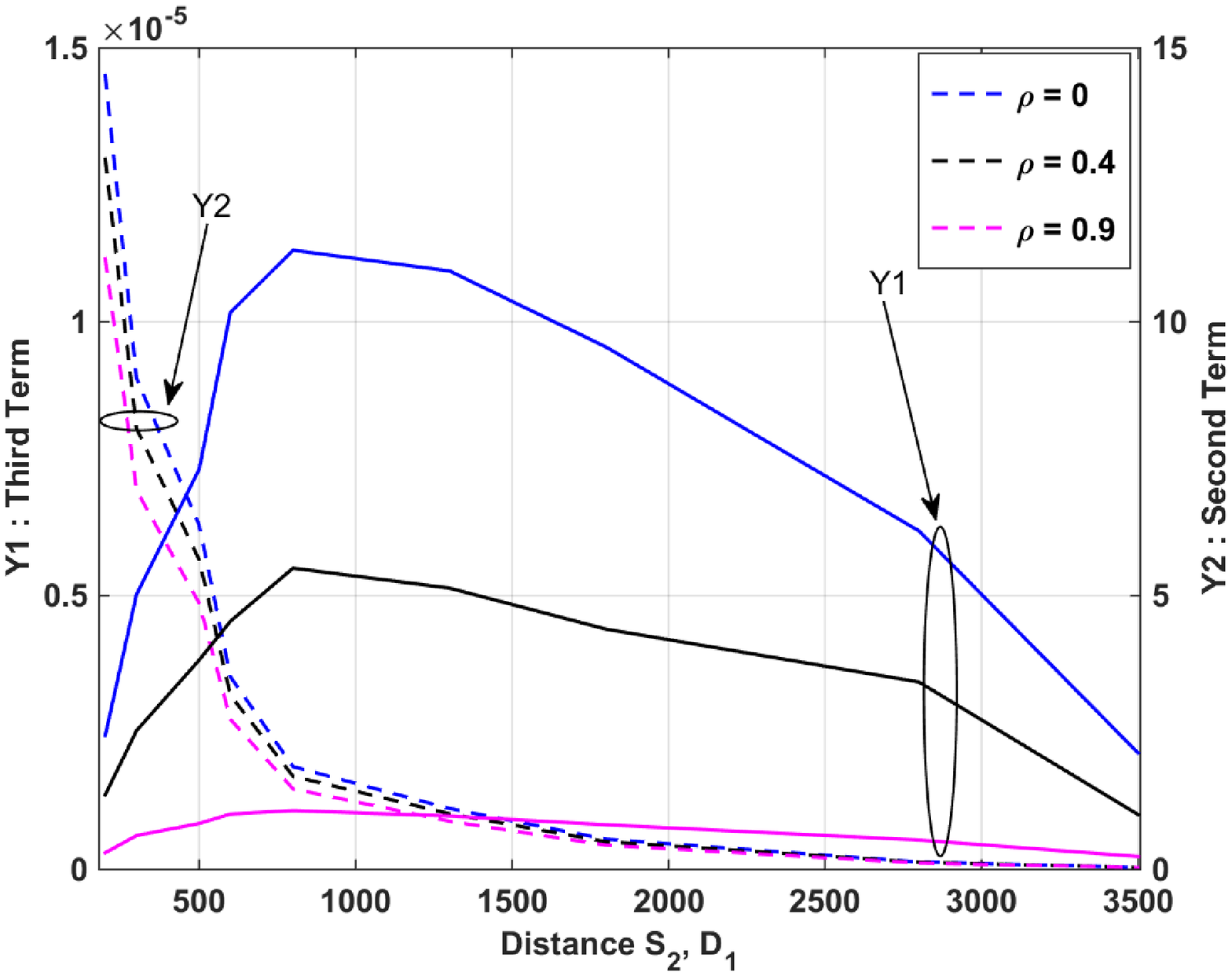}\\
  \caption{Variation of second and third term of (\ref{var2}) with distance.}
  \label{Fig:res6}
\end{figure}
\vspace{-2mm}

In order to explain this interesting behaviour of ANC noise variance with distance, we have plotted the second and third terms of expression of variance of ANC noise as given by (\ref{var2}) without considering the effect of $\sigma^2_o$ (Fig. \ref{Fig:res6}). In Fig. \ref{Fig:res6}, axes Y1 and Y2 show the third and second term, respectively. It can be noted that the behaviour of ANC noise variance observed in Fig. \ref{Fig:res4} and Fig. \ref{Fig:res5} is primarily governed by the third term. A careful look at the third term of (\ref{var2}) reveals that it represents the relative characteristics of the relay path to the direct path between $S_i$ and  $D_{j\neq i}$. Since, $d_{{S_2}R} < d_{{S_2}{D_1}}$, quality of received signal from $S_2$ is better at $R$ than $D_1$.
\begin{figure} [h!]
  \centering
  \includegraphics[height= 5.5 cm ,width=8.0cm]{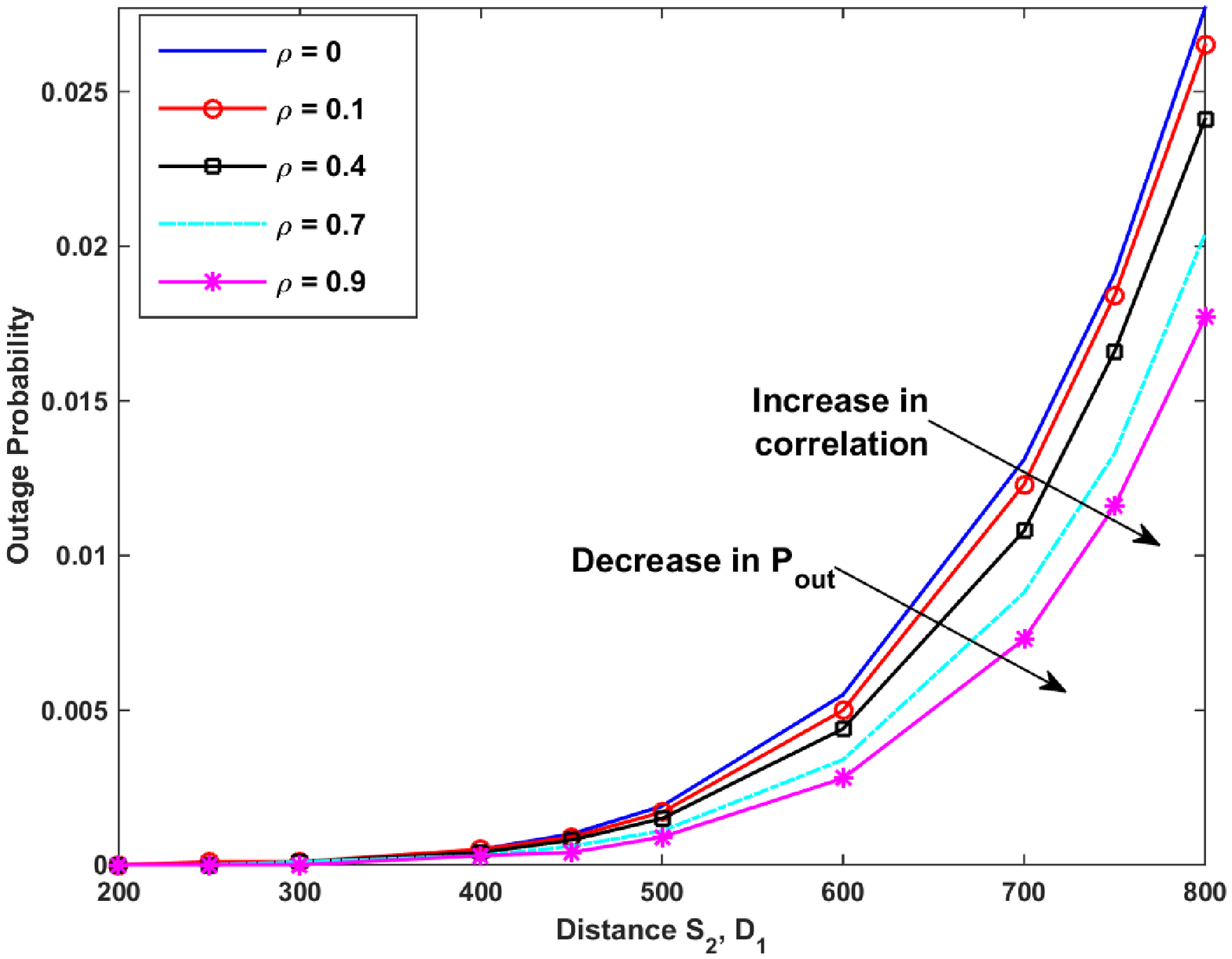}\\
  \caption{Outage probability with small distance for various correlation factors. }
  \label{Fig:res7}
\end{figure}
\vspace{-2mm}
This comparatively better signal is amplified by $R$ and forwarded to $D_1$. Though for smaller distance, signal quality at $D_1$, from both the paths (relay and direct) deteriorates but till the critical point rate of deterioration is less for relay path. Therefore, at any instance before the \emph{critical distance}, relayed signal remains stronger than the direct signal. Hence, subtraction of these two signals results in higher values of ANC noise variance. For larger values of $d_{S_2{R}}$, signal at $R$ also distorted enough due to higher pathloss. Further amplification of this (weaker) signal by $R$, results in a highly distorted signal at $D_1$. Therefore, After \emph{critical distance} signals from both the paths are considerably weaker and hence subtraction results in smaller values of variances.

To observe the effect of change of variance of ANC noise with correlation on the system performance, outage probability is calculated. Outage probability is defined as the probability that the received signal to noise ratio (SNR) falls below some certain threshold $\beta$. Numerically, outage probability can be found as \( P_\text{out} = P( \text{SNR}_\text{SC} < \beta) \), where $\text{SNR}_\text{SC}$ $= max\,[ \text{SNR}_\text{D}, \text{SNR}_\text{R} ]$. $\text{SNR}_\text{D}$ and $\text{SNR}_\text{R}$ are the SNR of direct path and relay path, respectively. If the received SNR falls below the threshold $\beta$, the bit is considered lost. Threshold $\beta$ is a dimensionless quantity which depends upon the specific application's QoS.
Outage performance of the network for smaller distance among various nodes for different correlation factors is shown in Fig. \ref{Fig:res7}.
An increase in distance between nodes ${S_2}$ and ${D_1}$ is associated to an  increase between ${S_1}$ and ${D_1}$ (direct path) or ${R}$ and ${D_1}$ (relay path) also, therefore, individual SNRs of both the paths go down.
Hence, $P_\text{out}$ increases with distance. However, $P_\text{out}$ decreases with an increase in correlation factor. This can be attributed to the outcomes of Fig. \ref{Fig:res4}. As the $\sigma _{\mathbb{N}_{{D_1}}^{NC}}^2\,$ decreases with $\rho$, noise power reduces in the signal from relay path, which inturn increases the SNR of the relay signal. This high SNR decreases the probability of outage event at $D_1$. This behaviour can be observed in  Fig. \ref{Fig:res7}.
\begin{figure} [ht!]
  \centering
  \includegraphics[height= 5.5 cm ,width=8.0cm]{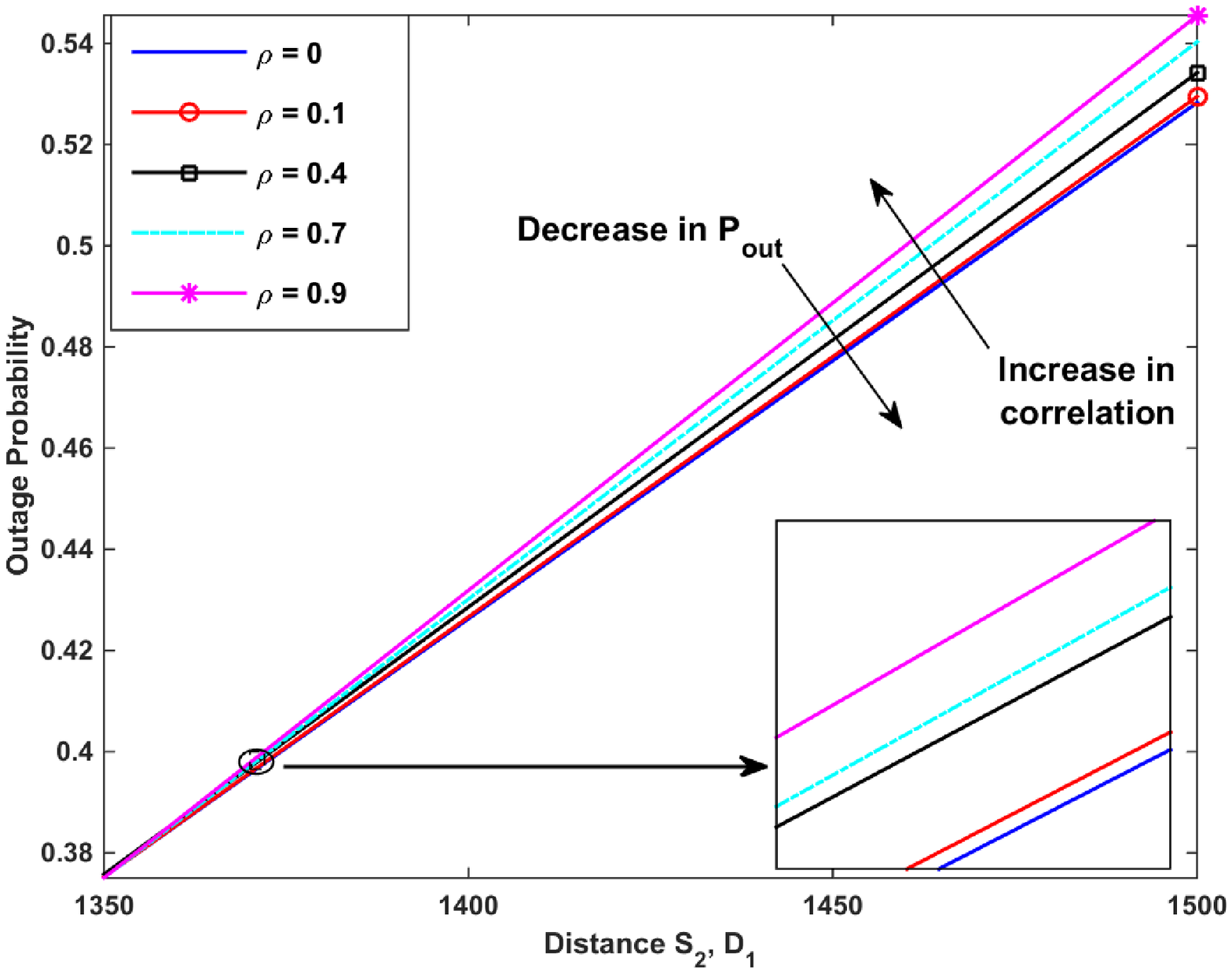}\\
  \caption{Outage probability with large distance for various correlation factors.}
  \label{Fig:res8}
\end{figure}
\vspace{-2mm}

In case of $ P_\text{out}$ also, a behaviour different from  Fig. \ref{Fig:res7} can be observed in Fig. \ref{Fig:res8} for larger distances. Unlike for smaller distances, after a certain point, outage performance degrades with increase in correlation.
A portion of curves after crossover point is shown separately in an additional box in Fig. \ref{Fig:res8}. It may be noted that after crossover point outage performance of different correlation scenarios is exactly in reverse order as compared to the order observed in Fig. \ref{Fig:res7}.
The reason behind this behaviour is the fact that increase in correlation among $h_{{S_2}{D_1}}$, $h_{{S_2}{R}}$ and $h_{{R}{D_1}}$ increases the probability of $h_{{R}{D_1}}$ being degraded with larger distance more than the uncorrelated case. Since, $h_{{R}{D_1}}$ is a part of $S_1-R-D_1$, degradation of  $h_{{R}{D_1}}$ results in lower values of SNR of signal from relay path at $D_1$. Therefore, after certain point, further increase in correlation results in higher outage at node $D_1$ as observed in Fig. \ref{Fig:res8}.



\label{sec:perf}
\section{Conclusion}

In this paper, we analyzed the effects of spatial correlation of different channels on ANC Noise at the receiver. It has been observed that ANC noise has deleterious effects on the system performance while its impact reduces with the increase in the spatial correlation among channels. The results are also presented for the characterization of the ANC noise with distance. The effect on network performance in terms of outage probability has been discussed as a function of distance and the channel correlation. Our work in this paper provides deep insight into the ANC noise characterization and outage probability of the system, which can be used by the network planners to provide better system efficiency and QoS to the users.


On the other hand, it should be noted that each destination node has to keep their receiver in active state to overhear the transmissions from other transmitters also. This increases the overall energy consumption of the system as compared to the case of no network coding at the relay. Therefore, scaling of the system is also limited by per node power constraint. Development of analytical framework for multiuser scenario to get better understanding of network coded cooperative scheme, behaviour of ANC noise with correlation and analysis of other asymmetrical topologies is considered as future work.

\label{sec:conc}

\bibliographystyle{IEEEtran}
\bibliography{IEEEabrv,References/ref}
\end{document}